\documentclass[aps,prl,twocolumn,showpacs,floatfix,altaffilletter, include graphics, longbibliography,superscriptaddress]{revtex4-2}
\usepackage{graphicx}
\usepackage{grffile}
\usepackage{amsmath}
\usepackage{amssymb}
\usepackage{bm}
\usepackage{color}
\usepackage{multirow}
\usepackage{tabularray}
\usepackage{booktabs}
\usepackage[dvipsnames]{xcolor}
\usepackage{amsmath,amssymb,amsfonts}
\usepackage{epsfig}
\usepackage{times}
\usepackage[colorlinks,bookmarks=false,citecolor=blue,linkcolor=blue,urlcolor=blue]{hyperref}
\usepackage[toc,page]{appendix}
\usepackage{float}
\usepackage{siunitx}
\usepackage{braket}
\usepackage{graphicx}
\usepackage[normalem]{ulem}

\newcommand{\tref}[1]{Table~\ref{#1}}

\newcommand{\msc}{m_{\rm SC}}
\newcommand{\dapi}{\delta_{\rm api}}
\newcommand{\vk}{\mathbf{k}}
\newcommand{\expp}[1]{\mathrm{e}^{#1}}
\newcommand{\cc}[2]{c_{#1}^{#2\phantom{\dagger}}}
\newcommand{\cdag}[2]{c_{#1}^{#2\dagger}}

\begin{document}

    %%%% TITLE %%%%%%%%%%%%%%%%%%%%%%%%%%%%%%%%%%%%%%%%%%%%%%%%%%%%%%%%%%%%%%%%%%%%
    \title{The Role of the Apical Oxygen in Cuprate High-Temperature Superconductors}
    %%%%%%%%%%%%%%%%%%%%%%%%%%%%%%%%%%%%%%%%%%%%%%%%%%%%%%%%%%%%%%%%%%%%%%%%%%%%%%%

    %%%% AUTHORS %%%%%%%%%%%%%%%%%%%%%%%%%%%%%%%%%%%%%%%%%%%%%%%%%%%%%%%%%%%%%%%%%%
    \author{Samuel Vadnais}
    \email{samuel.vadnais@utoronto.ca}
    \affiliation{Department of Physics, University of Toronto, Toronto, Ontario, Canada, M5S 1A1}
    \affiliation{D\'epartement de physique, Regroupement qu\'eb\'ecois sur les mat\'eriaux de pointe $\&$ Institut quantique Universit\'e de Sherbrooke, 2500 Boul. Universit\'e, Sherbrooke, Qu\'ebec J1K2R1, Canada}
    \author{Rémi Duchesne}
    \affiliation{D\'epartement de physique, Regroupement qu\'eb\'ecois sur les mat\'eriaux de pointe $\&$ Institut quantique Universit\'e de Sherbrooke, 2500 Boul. Universit\'e, Sherbrooke, Qu\'ebec J1K2R1, Canada}
    \author{Kristjan Haule}
    \affiliation{Center for Materials Theory, Department of Physics $\&$ Astronomy, Rutgers University, Piscataway, New Jersey 08854, USA}
    \author{A.-M. S. Tremblay}
    \affiliation{D\'epartement de physique, Regroupement qu\'eb\'ecois sur les mat\'eriaux de pointe $\&$ Institut quantique Universit\'e de Sherbrooke, 2500 Boul. Universit\'e, Sherbrooke, Qu\'ebec J1K2R1, Canada}
    \author{David S\'en\'echal}
    \affiliation{D\'epartement de physique, Regroupement qu\'eb\'ecois sur les mat\'eriaux de pointe $\&$ Institut quantique Universit\'e de Sherbrooke, 2500 Boul. Universit\'e, Sherbrooke, Qu\'ebec J1K2R1, Canada}
    \author{Benjamin Bacq-Labreuil}
    \email{benjamin.bacq-labreuil@ipcms.unistra.fr}
    \affiliation{Universit\'e de Strasbourg, CNRS, Institut de Physique et Chimie des Mat\'eriaux de Strasbourg, UMR 7504, F-67000 Strasbourg, France}
    \affiliation{D\'epartement de physique, Regroupement qu\'eb\'ecois sur les mat\'eriaux de pointe $\&$ Institut quantique Universit\'e de Sherbrooke, 2500 Boul. Universit\'e, Sherbrooke, Qu\'ebec J1K2R1, Canada}
    %%%%%%%%%%%%%%%%%%%%%%%%%%%%%%%%%%%%%%%%%%%%%%%%%%%%%%%%%%%%%%%%%%%%%%%%%%%%%%%

    \date{\today}

    %%%% ABSTRACT %%%%%%%%%%%%%%%%%%%%%%%%%%%%%%%%%%%%%%%%%%%%%%%%%%%%%%%%%%%%%%%%%
    
    \begin{abstract}
    Scanning tunneling microscopy measurements exploiting the natural superstructure modulation of the cuprate superconductor Bi$_2$Sr$_2$CaCu$_2$O$_{8+\delta}$ (Bi-2212) have revealed a possible correlation between the Cu-apical-O distance $\dapi$ and the superconducting order parameter $\msc$, as reported recently by O’Mahony et al. (Proc. Natl. Acad. Sci. 119, e2207449119 (2022)). 
    These observations were interpreted as evidence for a direct link between superconductivity and the charge-transfer gap, and more broadly revived the long-standing question of the role of apical oxygens in  cuprate superconductivity.
    Using a combination of density-functional theory and cluster dynamical mean-field theory, we compute from first principles the variations of $\msc$ induced solely by apical oxygen displacement in Bi$_2$Sr$_2$CuO$_{6+\delta}$, Bi-2212, and HgBa$_2$CuO$_{4+\delta}$. 
    The quantitative agreement between our calculations and experiments allows us to unambiguously attribute the observed variations of $\msc$ to changes in $\dapi$.
    We demonstrate, however, that these  variations of $\msc$ originate predominantly from changes in the effective hole-doping of the CuO$_2$ planes, with negligible effect  on the charge-transfer gap.
    The modest magnitude of the $\msc$ modulation induced by apical-oxygen displacement alone warrants caution in interpreting correlations between $T_c$ and $\dapi$ inferred from comparisons across different cuprate compounds.
    Our work demonstrates that the present \emph{ab initio} framework can quantitatively resolve the influence of specific structural degrees of freedom on superconductivity in correlated oxides.

\end{abstract}
    %%%%%%%%%%%%%%%%%%%%%%%%%%%%%%%%%%%%%%%%%%%%%%%%%%%%%%%%%%%%%%%%%%%%%%%%%%%%%%%
    
    \maketitle

%%%% MAIN TEXT %%%%%%%%%%%%%%%%%%%%%%%%%%%%%%%%%%%%%%%%%%%%%%%%%%%%%%%%%%%%%%%%%

%%%%%% Introduction
\paragraph{Introduction}

Considerable progress has been made in identifying physical parameters that optimize the superconducting transition temperature ($T_c$) of cuprate superconductors.
In particular, several experimental correlations have been found between increasing $T_c$ and (i) maximizing oxygen hole content~\cite{zheng1995,haase2004,jurkutat2014,rybicki2016,jurkutat2023}, (ii) stronger superexchange interaction~\cite{wang_paramagnons_2022}, and (iii) reduced charge transfer gap~\cite{ruan2016,wang_correlating_2023}. 
These correlations have been reproduced simultaneously in model calculations~\cite{kowalski2021}.
At the same time, advances in first-principles–based approaches~\cite{weber2012,cui_ab_2025,bacq-labreuil_towards_2025} raise the prospect that the long-standing problem of high-$T_c$ superconductivity in cuprates may be addressed quantitatively. 

Despite this progress, important questions remain unresolved, including the long-standing issue of the relationship between the apical oxygen distance ($\dapi$) and the superconductivity in cuprates~\cite{feiner1992,pavarini_band-structure_2001,kent2008,yin2009,weber2012,yee2014,peng2017,acharya2018}. 
Such an open problem represent a stringent test for the predictive capabilities of modern \emph{ab initio} approaches to high-temperature superconductors. 

Here, we take on this challenge motivated by an unexpected experimental observation.
A striking scanning tunnelling microscopy (STM) study of optimally-doped Bi$_2$Sr$_2$CaCu$_2$O$_{8+\delta}$ (Bi-2212)  revealed a strong correlation between the well-known super-structure modulation and the density of Cooper pairs, proportional to the squared superconducting order parameter $|\msc|^2$~\cite{omahony_electron_2022}.
The authors proposed that the superstructure modulation can be primarily understood as a modulation of $\dapi$, which in turn induces a spatial variation in the charge-transfer gap (CTG). 
This interpretation relies on the assumption that increasing $\dapi$ reduces the CTG~\cite{omahony_electron_2022}, and thereby enhances $\msc$~\cite{weber2012,wang_correlating_2023}.

Numerous studies have proposed correlations between the apical oxygen distance $\dapi$ and $T_c$ by comparing different cuprate families~\cite{pavarini_band-structure_2001,kent2008,weber2012,peng2017}.
These approaches implicitly assume that complex chemical and structural differences across materials can be effectively reduced to variations of a single parameter, $\dapi$, leaving unresolved the question of whether there exists a direct causal link with $T_c$.
A few works attempted to isolate the specific role of $\dapi$. 
Feiner~\textit{et al.} studied the coupling to out-of-plane degrees of freedom within a five-band Hubbard model~\cite{feiner1992}, focusing in particular on how 
apical oxygen hybridization affects the contribution of
Cu-$d_{z^2}$-like orbitals to the low-energy Zhang-Rice (ZR)~\cite{zhang1988} band. 
This line of reasoning connects naturally to more recent studies which, without explicitly considering the apical oxygen, argue that pushing the Cu-$d_{z^2}$ orbital away from the Fermi level is favorable for superconductivity~\cite{sakakibara2010,sakakibara2012}.
The isolated effect of apical oxygen on the parameters of low-energy effective models for cuprates has also been studied in Refs.~\onlinecite{yin2009,yee2014}.
Nevertheless, as in Ref.~\onlinecite{feiner1992}, a direct impact on the superconducting properties could not be established. 
Finally, a more recent study by Acharya \textit{et al.} employed advanced \textit{ab initio} methods to relate controlled displacements of the apical oxygen in La-based cuprates to changes in $\msc$~\cite{acharya2018}. 
However, $\msc$  was assessed without explicitly considering hole-doping; strong correlations were treated at the level of single-site DMFT; and superconductivity was inferred using a Bardeen–Cooper–Schrieffer–based ansatz, whose applicability to cuprates is questionable.

Here, we address three crucial questions that have remained unanswered to date: 
(i) Can the effect of apical oxygen displacements on the superconducting properties of cuprates be quantified?
(ii) Are the apical oxygen displacements alone sufficient to account for the experimentally observed periodic variations of $\msc$ in Bi-based cuprates?
(iii) Does the charge-transfer gap mediate the relationship between $\dapi$ and $\msc$?

We answer these three questions using a recently developed \textit{ab initio} framework~\cite{bacq-labreuil_towards_2025} that enables first-principles predictions of the superconducting order parameter $\msc$ by combining density functional theory (DFT)~\cite{hohenberg1964,kohn1965} with cluster dynamical mean-field theory (CDMFT)~\cite{lichtenstein1998,lichtenstein2000,kotliar2001,kotliar2006}. 
By performing calculations on single-layer Bi$_2$Sr$_2$CuO$_{6+\delta}$ (Bi-2201) and bi-layer Bi-2212 in which $\dapi$ is varied, we quantitatively reproduce the experimental observations of O'Mahony~\textit{et al.}~\cite{omahony_electron_2022}. 
Our approach enables us to assign an absolute scale to the experimentally measured relative variations of $|\msc|^2$, and to show that the effect of $\dapi$ is small compared to other changes in the crystal structure. 
This implies that the proposed correlation between $\dapi$ and $T_c$ inferred from comparisons across different cuprate families~\cite{pavarini_band-structure_2001,kent2008,weber2012,peng2017} 
should be treated with caution.
Finally, our calculations reveal that variations of $\msc$ with $\dapi$ cannot be explained by changes in the CTG.
Instead, we find that apical oxygen displacements primarily act by modulating the \emph{effective} hole doping of the CuO$_2$ planes, involving the Cu-$d_{x^2-y^2}$ and O-$p_{x/y}$ orbitals, which in turn controls $\msc$.
We perform additional calculations for HgBa$_2$CuO$_{4+\delta}$ (Hg-1201), which further support this interpretation.
Our work establishes that variations in the \emph{effective} hole doping of the CuO$_2$ planes play a central role and must be carefully accounted for.

\begin{figure}
\includegraphics[width=1.0\linewidth]{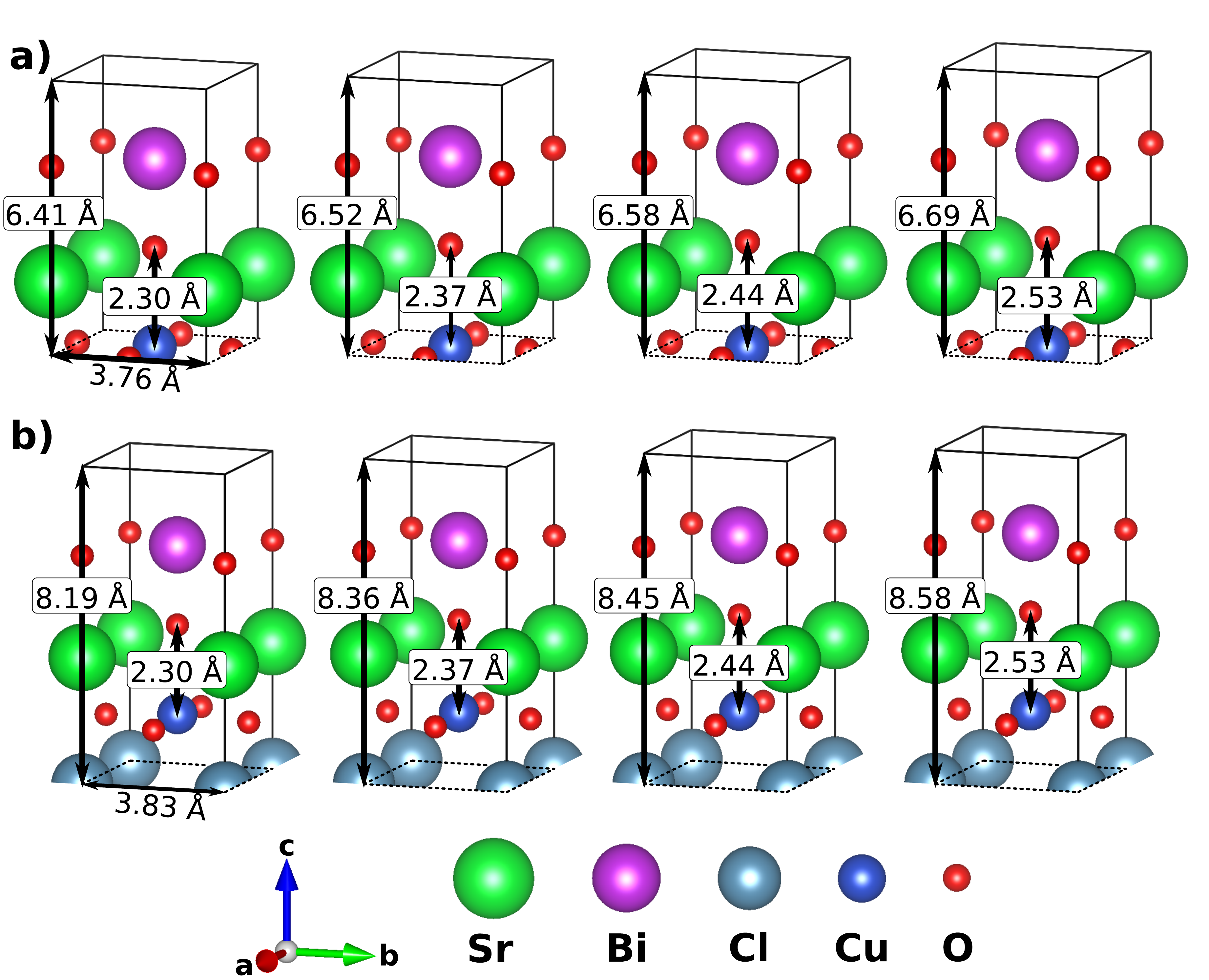}
    \caption{Half simplified ($P4/mmm$) unit cells of (a) Bi-2201 and (b) Bi-2212. 
    $\delta_{\rm api}$, $c/2$ and $a$ are displayed (See SM~\cite{SupMat} for details). 
\label{fig:fig1}
    }
\end{figure}
%

%%%%%% Method
\paragraph{Method}

A fully first-principles description of  $\msc$ in Bi-2201 and Bi-2212 that explicitly incorporates the long-range superstructure modulation is currently intractable.
STM measurements show that this modulation produces stripe-like regions, in which the tetragonal crystal structure is preserved, along the diagonals of the CuO$_2$ planes with respect to the $a$ and $b$ axes~\cite{mcelroy2005,slezak2008,omahony_electron_2022}. %, as illustrated in Fig.~\ref{fig:fig1}.
The short superconducting coherence length in cuprates~\cite{hwang2021}, together with the success of previous small cluster DMFT calculations for superconductivity, suggests that 
the dominant contribution arises from local structural variations of $\dapi$.
Accordingly, we perform calculations for homogeneous Bi-2201/2212 structures with four distinct values of $\dapi$,  chosen to match the distances observed experimentally~\cite{omahony_electron_2022}, as illustrated in Fig.~\ref{fig:fig1}~\footnote{Model calculations based on an inhomogeneous Hubbard model with a modulation wavelength of 12-sites, close to experiment, lead to a correlation between hole-doping and $\epsilon_p-\epsilon_d$ modulation similar to what is found in our first-principles work, as can be seen in Fig.~C.0.13 of Ref.~\cite{rotella:tel-04995607}. In those calculations, $\epsilon_p-\epsilon_d$ is a phenomenological parameter.}.

Since atomic displacements involve energy scales much larger than those related to superconductivity, we optimized the Bi-2201 unit cell parameters ($a$, $c$) and internal atomic positions using the eDMFT approach~\cite{haule2010,haule2016,haule2018},  while keeping the apical oxygen distance $\dapi$ fixed.
Increasing $\dapi$ leads to an expansion of the $c$ lattice parameter, while $a$ remains essentially unchanged.
The Bi-2212 structures were constructed empirically by assuming that the dependence of $a$ and $c$ on $\dapi$ follows the same trends as in Bi-2201.
The resulting cell parameters $a$, $c$, and $\dapi$ are summarized in Fig.~\ref{fig:fig1}.

We then compute $\msc$ in Bi-2201 and Bi-2212 using the DFT+CDMFT framework of Ref.~\onlinecite{bacq-labreuil_towards_2025}. 
We construct $2\times2$ supercells, with a correlated subspace restricted to four Cu-$d_{x^2-y^2}$ orbitals, and perform DFT+CDMFT calculations at effectively zero temperature with $U=9~\text{eV}$. 
Since Bi-2201 and Bi-2212 are naturally self-doped~\cite{lin2006,nokelainen2020}, no virtual crystal approximation (VCA)~\cite{bellaiche_virtual_2000} is required. 
These calculations employ the eDMFT package~\cite{haule2010,haule2016,haule2018,bacq-labreuil_towards_2025}, extending Wien2k~\cite{Wien2K} to DFT with (C)DMFT, and combined with the exact diagonalization solver implemented in PyQCM~\cite{PyQCM}.
The Wannier90 package~\cite{mostofi2008,mostofi2014} is used to construct maximally localized Wannier orbitals used in Fig.~\ref{fig:fig3} and End Matter (EM).
We refer to the SM for more computational details~\cite{SupMat} and to Ref.~\onlinecite{bacq-labreuil_towards_2025} for an in-depth presentation of our DFT+CDMFT framework.

%%%%%% Results
\paragraph{Results}

We first address the above questions (i) and (ii) by showing how $\msc$ varies with $\dapi$ in Bi-based cuprates. 
We then answer (iii) in two steps.
By downfolding to an effective three-band model, we demonstrate that the CTG \emph{increases} with increasing $\dapi$, thereby ruling out the scenario of Ref.~\onlinecite{omahony_electron_2022}.
This result leads to our central conclusion: the variations of pair density $|\msc|^{2}$ are instead governed by the change in effective doping induced by the apical oxygen displacement.
Finally, we substantiate this conclusion by extending our analysis to Hg-1201.

\begin{figure}
    \includegraphics[width=0.85\linewidth]{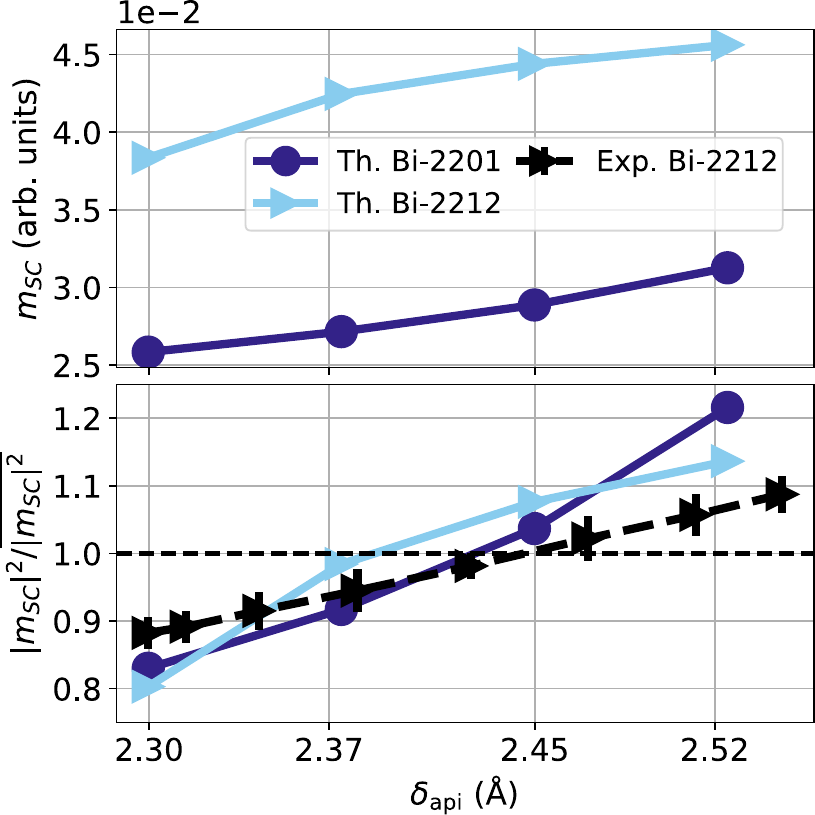}
        \caption{Computed superconducting order parameter $\msc$ vs $\dapi$ for Bi-2201 and Bi-2212 (top). 
        Note the scaling of the $y$ axis. 
        Relative variations of the computed and measured superfluid density $|\msc|^2/\overline{|\msc|^2}$ vs $\dapi$ (bottom).
        The experimental data are digitized from Ref.~\onlinecite{omahony_electron_2022}.
        }
    \label{fig:fig2}
\end{figure}

\begin{figure}
    \centering    
    \includegraphics[width=1.0\linewidth]{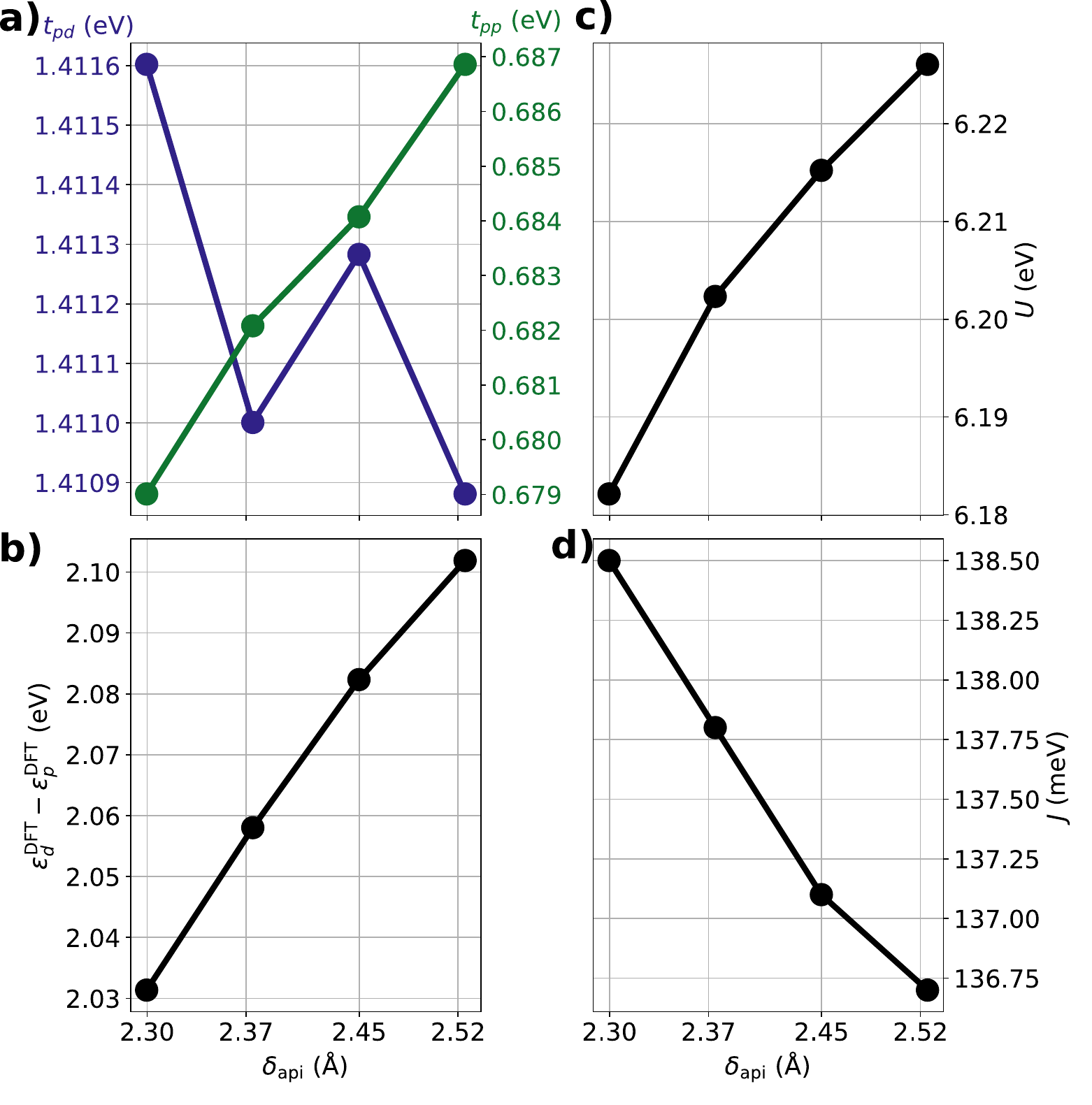}
        \caption{(a) $t_{pd}$ and $t_{pp}$ (Wannier), (b) $\left(\varepsilon^{DFT}_d-\varepsilon^{DFT}_p\right)$ (Wannier), (c) $U$ (constrained DFT) and (d) $J$ (Cu$_2$O$_{25}$ clusters) vs $\delta_{\rm api}$. 
        All parameters were computed from Bi-2201.}
    \label{fig:fig3}
\end{figure}

Figure~\ref{fig:fig2} presents the computed $m_{SC}$ in Bi-2201/2212. 
The absolute variations of $\msc$ with $\dapi$ are relatively weak, of order $0.5\times10^{-2}$ for both compounds, and approximately three times smaller than the difference between $\msc$ for Bi-2201 and Bi-2212. 
Note that $\msc$ of Bi-2201 is about twice that of Bi-2212, consistent with the measured $T_c$ of Bi-2212 being nearly twice that of Bi-2201~\cite{wang_paramagnons_2022,wang_correlating_2023}.

The relative variations of the superfluid density, $|\msc|^2/\overline{|\msc|^2}$ is of order $30\%$ and it shows remarkable quantitative agreement with the STM measurements reported in  Ref.~\onlinecite{omahony_electron_2022} on Bi-2212 (see Fig.~\ref{fig:fig2}).
The computed variations in Bi-2212 exhibit a slightly more pronounced curvature. 
We attribute this to the construction of the unit cells, in which the same $c$-axis versus $\dapi$ relation as in Bi-2201 was assumed.
This modeling choice does not affect our central conclusion: the apical oxygen displacement alone accounts for the measured variations of $|\msc|^2$ associated with superstructure modulations in Bi-2212, and is expected to apply similarly to Bi-2201~\cite{wang_correlating_2023}.
Most importantly, our analysis puts in perspective the variations induced by $\dapi$ and shows that they are small in comparison to the differences observed between distinct cuprate materials.
This observation calls for caution when interpreting the proposed correlation between $\dapi$ and $T_c$~\cite{pavarini_band-structure_2001,kent2008,weber2012,peng2017}.
%
%The ability of \emph{ab initio} approaches such as ours to quantify variations of the superconducting properties is expected to be key for guiding the design of novel high-temperature superconductors. 

Before examining whether the CTG could be the link between $\msc$ and $\dapi$, we first clarify its definition.  
The CTG corresponds to the energy required to transfer an electron from an O to a Cu atom. 
It is well defined in \emph{insulating} cuprates since it coincides with the spectral gap.
Indeed, the highest occupied O-$p$ states belong to the low-energy ZR band, and the lowest unoccupied Cu-$d$ states form the upper Hubbard band~\cite{bacq-labreuil_towards_2025}.
Upon hole-doping, strong electronic correlations are known to induce substantial spectral weight transfer of Cu-$d$ character to energies just above the Fermi level~\cite{meinders1993,bacq-labreuil_towards_2025}. 
A strict application of the above definition, therefore, implies that the CTG vanishes in hole-doped \emph{metallic} compounds.
To define in such a case an energy scale corresponding to the CTG, theoretical studies proposed to consider the energy separation between the upper edge of the ZR singlet (above the Fermi level) and the upper Hubbard band~\cite{kowalski2021}, or the difference between the on-site energies of the Cu-$d_{x^2-y^2}$ and O-$p_{x/y}$ orbitals~\cite{weber2012}.
All of these definitions associate variations of the CTG in hole-doped compounds with changes in the electronic structure \emph{above} the Fermi level, which itself remains fixed.
Crucially, none of these definitions corresponds to that employed in Ref.~\onlinecite{omahony_electron_2022}, where the proposed CTG variations are instead inferred from a redistribution of spectral weight \emph{below} the Fermi level.

Bi-based cuprates are subject to self-doping~\cite{lin2006}, preventing us from performing calculations in the insulating regime where the CTG is well defined. 
To circumvent this problem, we resort to an effective Emery--Varma--Schmitt-Rink--Abrahams model~\cite{emery1987,varma1987}.
Setting the Cu-$d_{x^2-y^2}$ on-site energy to zero, $\Delta_p$ denotes the on-site energy of the O-$p_{x/y}$ orbitals, $\mu$ is the chemical potential, and $t_{pd}$ ($t_{pp}$) are the Cu--O (nearest O-O) hopping amplitudes (see SM~\cite{SupMat}).
For each value of $\dapi$, these parameters are obtained by constructing maximally localized Wannier orbitals from a DFT-only calculation for Bi-2201.
The energy separation between Cu and O orbitals requires a double-counting correction, for which we adopt the nominal scheme~\cite{pourovskii2007,haule2010,haule2014}.
Assuming one electron per Cu-$d_{x^2-y^2}$ orbital,  $\Delta_p$ is then given by $\Delta_p = \epsilon^{\rm DFT}_{p} - \epsilon^{\rm DFT}_{d} + \frac{U}{2},$ where $\epsilon^{\rm DFT}_{p(d)}$ are the on-site energies of the Wannier orbitals.

We show in Fig.~\ref{fig:fig3}(a,b) $t_{pd}$, $t_{pp}$ and $(\epsilon^{\rm DFT}_{d} - \epsilon^{\rm DFT}_{p})$ with respect to $\dapi$. 
The hopping parameters remain essentially constant. 
By contrast, $(\epsilon^{\rm DFT}_{d} - \epsilon^{\rm DFT}_{p})$, which could be identified to the CTG at the DFT level~\cite{weber2012}, increases by a small but significant amount of approximately $\sim 3-4\%$, while $\dapi$ varies by about $10\%$. 
In the simplified perspective of the three-band model, the bare CTG is expected to scale as $\Delta_{\rm CTG} \simeq U - \Delta_{p} - W = \frac{U}{2} + (\epsilon^{\rm DFT}_{d} - \epsilon^{\rm DFT}_{p}) - W$, with $W$ the bandwidth. 
%From a simplified mean-field perspective, the bare CTG is expected to scale as $\Delta_{\rm CTG} \simeq U - \Delta_{p}$.
%
Since $(\epsilon^{\rm DFT}_{d} - \epsilon^{\rm DFT}_{p})$ is an increasing function $\dapi$, and the hopping amplitudes (thus $W$) are almost constant, $\Delta_{\rm CTG}$ is therefore expected to be an \emph{increasing} function of $\dapi$, in direct contradiction with the interpretation of Ref.~\onlinecite{omahony_electron_2022}.

To complete this argument, it is necessary to examine the $\dapi$ dependence of the on-site interaction $U$ on the Cu-$d_{x^2-y^2}$ orbitals.
We estimated $U$ using constrained DFT~\cite{anisimov1993} with $\sqrt{5}\times\sqrt{5}$ supercells. 
This approach is known to underestimate $U$ for cuprates, and the supercells we consider are modest. 
We argue that the systematic error is the same for all $\dapi$ for a given compound (the main structural change is $\dapi$), and thus that the trend is meaningful.
Hence, the values of $U$ found from cDFT should only be seen as a way to assess the relative variations of the on-site interaction with respect to $\dapi$. 
The $U=9~\unit{eV}$ used in the DFT+CDMFT is, instead, an effective on-site interaction appropriate for the cluster correlated subspace. 
As shown in Fig.~\ref{fig:fig3}(c), $U$ is found to be a \emph{slowly increasing} function of $\dapi$, reflecting the reduced screening on the Cu-$d$ orbitals as the apical oxygen moves away from the Cu site. 
Accounting for the $\dapi$ dependence of $U$ would therefore further reinforce our conclusion that the CTG is an increasing function of $\dapi$.

To further substantiate our interpretation, we compute the superexchange interaction $J$ of the 3-band model using a more appropriate $U=9~\text{eV}$ (same as in the DFT+CDMFT calculations) and exact diagonalization of Cu$_2$O$_{25}$ clusters~\cite{eskes1993}.
The energy difference between the lowest-energy singlet and triplet states provides an estimate of $J$. 
The resulting values are close to the experimentally reported estimates ($\sim 150~\text{meV}$~\cite{peng2017,wang_paramagnons_2022}) and most importantly, $J$ is found to be a decreasing function of $\dapi$ (see Fig.~\ref{fig:fig3}(d)). 
Since $J\propto t_{\mathrm{eff}}^{2}/\Delta_{\mathrm{CTG}}$, the observed decrease of $J$ is internally consistent.
A decreasing $J$ is incompatible with the increase of $\msc$~\cite{wang_paramagnons_2022}, and therefore the CTG scenario proposed in Ref.~\onlinecite{omahony_electron_2022} is not viable.

Although our analysis focuses on Bi-2201 for convenience, the conclusions directly extend to Bi-2212. 
Moreover, we performed direct calculations of the CTG as a function of $\dapi$ in insulating Hg-1201, as presented in the SM~\cite{SupMat}. 
The results confirm the above interpretation, and show that the same behaviour is expected for other cuprates.

Our analysis calls for a new interpretation of the $dI/dV$ measurements of Ref.~\onlinecite{omahony_electron_2022} (see Fig.~3B therein).
As discussed above, variations of the CTG at constant hole doping should manifest as changes in the \emph{unoccupied} part of the electronic spectrum. 
Instead, the $dI/dV$ measurements reveal variations of the spectral weight predominantly in the \emph{occupied} part of the spectrum~\cite{omahony_electron_2022}, consistent with changes in the effective occupation of the CuO$_2$ planes.
 
\begin{figure}
    \centering
    \includegraphics[width=0.9\linewidth]{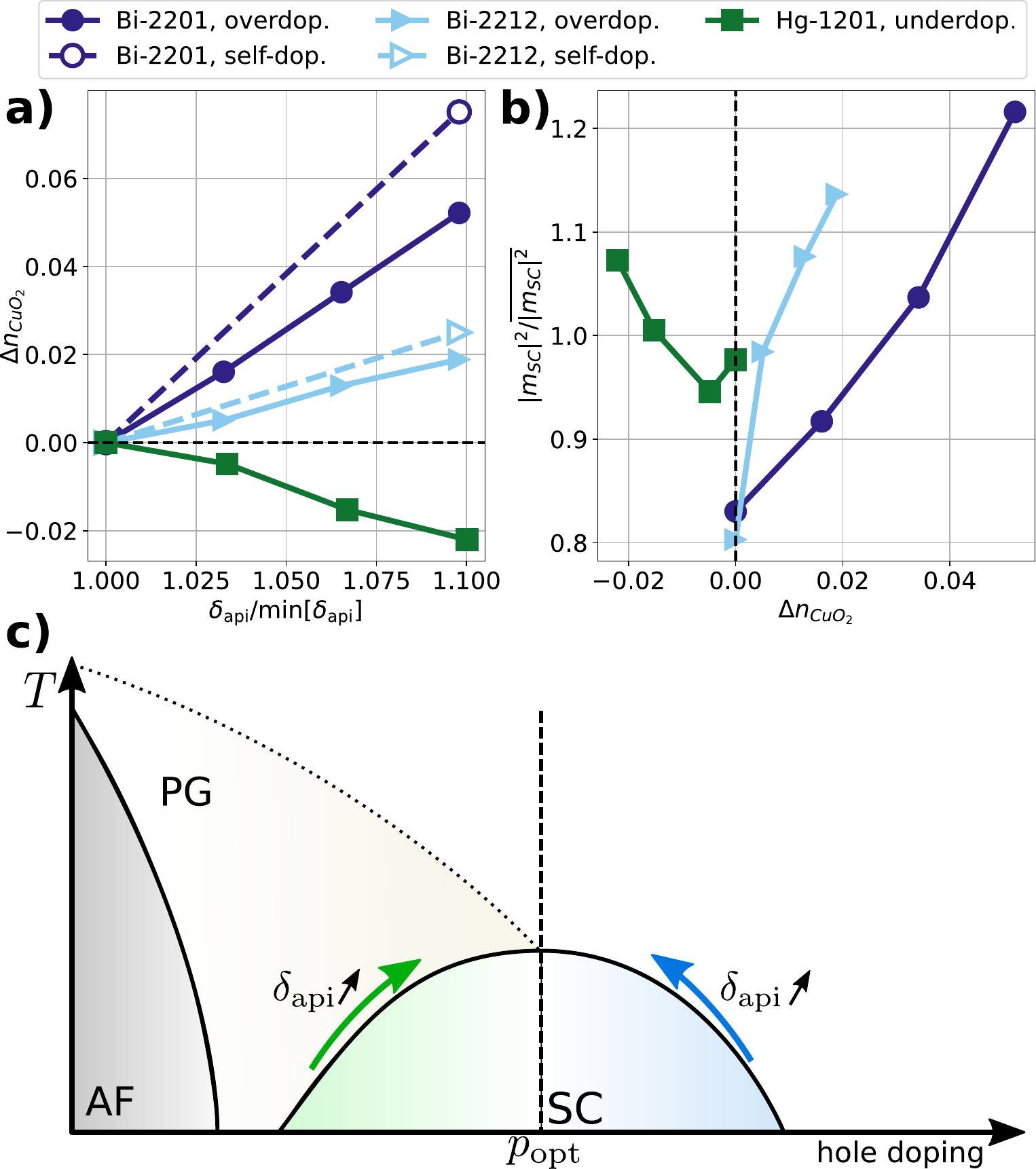}
    \caption{(a) DFT+CDMFT electron occupation variations $\Delta n_{\rm CuO_2}$ of the in-plane Cu-$d_{x^2-y^2}$ and O-$p_x/p_y$ orbitals (solid lines, filled symbols) vs $\delta_{\rm api}/{\rm min}\left[\delta_{\rm api}\right]$.
	The dashed lines with open symbols represent the expected occupation change obtained from the wannierization of the self-doping bands in Bi-2201 and Bi-2212 (see text and SM~\cite{SupMat} for details). 
    (b) Relative variations of the superfluid density $|\msc|^2/\overline{|\msc|^2}$ vs $\Delta n_{\rm CuO_2}$. 
    (c) Representative sketch of the cuprate's phase diagram.
    $m_{SC}$ is an increasing function of $\dapi$ in all compounds since the effective hole doping increases in underdoped Hg-1201 (green), while it decreases in overdoped Bi-2201 and Bi-2212 (blue). 
    }
    \label{fig:fig4}
\end{figure}

We therefore analyze the electron occupation of the CuO$_2$ planes obtained from our \textit{ab initio} DFT+CDMFT calculations for Bi-2201 and Bi-2212. 
We consider the variations of the occupation of the Cu-$d_{x^2-y^2}$ and O-$p_{x/y}$ orbitals, $\Delta n_{\rm CuO_2}$, around the value obtained for the minimal $\dapi$, as shown in Fig.~\ref{fig:fig4}(a).
In both Bi-2201 and Bi-2212, we observe an increase in electron occupation, corresponding to a reduction of the effective hole doping with increasing $\dapi$.
Because self-doping places the CuO$_2$ planes in the overdoped regime (see EM), increasing $\dapi$ drives the CuO$_2$ planes closer to optimal doping.
This, in turn, leads to an enhancement of $\msc$ (see Fig.~\ref{fig:fig4}(b,c)) reflecting its strong sensitivity to doping.
Quantitatively, variations of $\delta_{api}$ by $\sim 10\%$ induces changes in the effective doping of about $\sim 5\%$ ($\sim 2\%$ per CuO$_2$ plane) in Bi-2201 (Bi-2212).
Despite the weaker per-plane doping variations in the bi-layer compound, the relative changes in $\msc$ are comparable to those observed in single-layer Bi-2201.

The same analysis is repeated for Hg-1201, for which calculations can be carried out in the underdoped regime. 
We fix a total doping of $15\%$ using VCA~\cite{bellaiche_virtual_2000}, corresponding to an effective doping of approximately $\sim 11-12\%$ in the CuO$_2$ planes~\cite{bacq-labreuil_towards_2025}. 
Hg-1201 structures were constructed in a similar way to Bi-2212, 
except that the apical oxygen distances are systematically larger in Hg-1201 (see SM~\cite{SupMat}): $\dapi^{\rm Hg}\in\left[2.56,2.81\right]\mathrm{\AA}$, $\dapi^{\rm Bi}\in\left[2.3,2.52\right]\mathrm{\AA}$. 
Remarkably, $\Delta n_{\rm CuO_2}$ is found to anti-correlate with $\dapi$. Since Hg-1201 is here underdoped, we find that $\msc$ again increases with increasing $\dapi$, as shown in Fig.~\ref{fig:fig4}(a,b). 
These results are schematically summarized in Fig.~\ref{fig:fig4}(c).
A detailed analysis of the $\dapi$ dependence of $\msc$ in Hg-1201 is provided in SM~\cite{SupMat}.

It is remarkable that increasing $\dapi$ leads to opposite trends in $\Delta n_{\rm CuO_2}$ in Hg-1201, compared to Bi-2201/2212. 
It deserves an analysis. 
We find it fruitful to think of the charge transfer processes in terms of distinct \emph{channels}. 
All compounds considered here share a first common channel:  achieving the favorable oxidation state of all atoms requires the charge reservoir layers to \emph{donate} electrons to the nominally doped Cu$^{2+}$(O$^{2-}$)$_2$ planes globally. 
Increasing $\dapi$ weakens this channel in \emph{all compounds} because it has the same physical origin, thereby leading to \emph{increased} effective hole-doping, as in Hg-1201. 
For Bi-2201/2212, however, a second channel is present, associated with the Bi-O self-doping bands, which promote charge transfer in the opposite direction -- from the CuO$_2$ planes to the charge reservoirs. 
Weakening this channel, therefore, results in a \emph{reduction} of the effective hole doping.
Because this second channel involves the hybridization of the CuO$_2$ orbitals with only a small number of self-doping bands, it is expected to be more sensitive to $\dapi$ than the first channel, which reflects global oxidation constraints and does not rely on a limited set of bands.

Our hypothesis can be quantified: based on DFT-only calculations we evaluated the effective electron occupation of the self-doping bands in Bi-2201 and Bi-2212 for the lowest ($\dapi=2.30~\text{\AA}$) and largest ($\dapi=2.52~\text{\AA}$) apical distances.
We wannierized the self-doping bands, computed the corresponding density of states and extracted an effective occupation change $\Delta n_{\rm CuO_{2}}$, as explained in SM~\cite{SupMat}. 
As displayed in Fig.~\ref{fig:fig4}(a) (dashed lines, empty symbols), for both Bi-2201 and Bi-2212 the self-doping contribution to $\Delta n_{\rm CuO_{2}}$ exceeds the DFT+CDMFT estimation. 
Interestingly, the difference is roughly $0.02~\mathrm{e}^{-}$ in Bi-2201, which corresponds well to the decrease in occupation of roughly $0.022~\mathrm{e}^{-}$  for Hg-1201. 
This is consistent with the existence of the competing channel common to Bi- and Hg-based compounds. 
The per-plane difference between the self-doping channel and the DFT+CDMFT estimate is smaller in Bi-2212, suggesting that the common global charge transfer channel is less affected by $\dapi$ in bi-layer systems.
While a direct estimation of this common channel could be envisaged, this indirect approach is sufficient to support our interpretation. 

%%%%%% Discussion
\paragraph{Discussion}

We have presented material-specific predictions that clarify the role of the apical oxygen in the superconducting properties of cuprates. 
By answering the three central questions raised in introduction, we not only quantify the effect of the apical oxygen distance $\dapi$ on the superconducting properties, but we also elucidate the underlying mechanism. 
This constitutes a demonstration that \emph{ab initio} approaches like ours enable the prediction of accurate and quantified guidelines to stimulate the search for novel high-temperature superconductors.

Variations of $\dapi$ of the order of $10\%$ lead to relative changes of $\msc$ of approximately $20\%$, and to relative variations of the superfluid density of about $15\%$ in Hg-1201, $30\%$ in Bi-2212, and $40\%$ in Bi-2201.
We also provided an absolute scale to variations in $\msc$ that are modest compared to those induced by other structural modifications, such as the transition from single-layer to bi-layer cuprates [question (i)].
This observation calls for caution when interpreting the proposed correlations between $\dapi$ and $T_c$ inferred from comparisons across different cuprate families~\cite{pavarini_band-structure_2001,kent2008,weber2012,peng2017}.

The quantitative agreement between our calculations and the STM measurements of Ref.~\cite{omahony_electron_2022} establishes apical oxygen displacement as the relevant microscopic mechanism underlying the periodic modulations of the superconducting order parameter in Bi-based cuprates [question (ii)].
At the same time, we provide compelling evidence, from the study of three different compounds, that the variations of $\msc$ induced by $\dapi$ are not governed by changes in the CTG.
While the CTG remains a key energy scale for superconductivity in cuprates, it does not control the effect studied here.
Instead, we demonstrate that variations of $\dapi$ modify the effective hole-doping of the CuO$_2$ planes, which in turn drives the observed changes in $\msc$ [question (iii)]. 
We emphasize that our findings are further supported by experimental evidence provided in Ref.~\cite{antipov2002}. 
Samples of Hg-1201 have been hole-doped in two ways, inserting O (formal valence O$^{2-}$) or inserting F (formal valence F$^{-}$).
At optimal doping, the amount of F dopants is nearly twice that of O, as expected from the formal valence, yet the $T_c$ and the lattice constant $a$ are the same. 
The main difference between the F- and O-doped compounds is the apical distance $\dapi$, which is smaller in the F-doped samples~\cite{antipov2002}. 
This demonstrates that, at \emph{fixed doping}, the influence of $\dapi$ on $T_c$ is small, if not negligible, thus confirming our interpretation. 

%%%%%% Conclusion
\paragraph{Conclusion}
It has been established long ago that doping plays a crucial role in cuprates and other metal-oxide superconductors. 
The recently developed correlated electronic structure framework employed here has recently demonstrated~\cite{bacq-labreuil_towards_2025} that layer-dependent doping leads to natural explanations of many paradoxes encountered in multilayer cuprates.
Here we go one step further, establishing the role of specific atomic positions in determining doping and superconducting tendencies of several compounds: overdoped Bi-2201 and Bi-2212, and underdoped Hg-1201.
The interpretation in terms of doping channels motivates further investigations to achieve an optimal control of doping in cuprates, as well as nickelates~\cite{dicataldo2024}.
More broadly, our work opens the way to systematic investigations of how the chemical composition and the atomic positions may affect $T_c$ in cuprates. 
Open questions, such as the impact of the oxygen octahedra tilting, the role of atomic positions under pressure, and lack of apical oxygen in certain compounds, to name a few, can now be investigated.
Addressing these questions may be important for guiding the design of improved high-temperature superconductors.

\paragraph{Data availability}

The data and scripts used to generate the figures, as well as the cif structure files have been deposited in the Open Science Framework repository~\cite{OSF-repo}. 

\paragraph{Acknowledgments}
We are grateful to Marcello Civelli for discussions on model calculations for inhomogeneous lattices~\cite{rotella:tel-04995607}.
This work has been supported by the Natural Sciences and Engineering Research Council of Canada (NSERC) under grant ALLRP~588280-23 and by the Canada First Research Excellence Fund. 
The Canadian Foundation for Innovation, the Ministère de l'Économie, de l'Innovation et de l'Énergie (Québec), Calcul-Québec and the Digital Research Alliance of Canada provided part of the computational resources.
This project was provided with computing HPC and storage resources by GENCI at TGCC, thanks to the grant 2024-A0170915694 on the supercomputer Joliot Curie's SKYLAKE and ROME partitions. 
This work of the Interdisciplinary Thematic Institute QMat, as part of the ITI 2021-2028 program of the University of Strasbourg, CNRS and Inserm, was supported by IdEx Unistra (ANR 10 IDEX 0002), and by SFRI STRAT'US project (ANR 20 SFRI 0012) and EUR QMAT ANR-17-EURE-0024 under the framework of the French Investments for the Future Program. KH acknowledges support from NSF DMR-2233892 and a grant from the Simons Foundation (SFI-MPS-NFS- 00006741-06).
D.S. and A.-M.S. benefit from their RQMP membership (https://doi.org/10.69777/309032).
%%%%%%%%%%%%%%%%%%%%%%%%%%%%%%%%%%%%%%%%%%%%%%%%%%%%%%%%%%%%%%%%%%%%%%%%%%%%%%%

%%%%% BIBLIOGRAHPY %%%%%%%%%%%%%%%%%%%%%%%%%%%%%%%%%%%%%%%%%%%%%%%%%%%%%%%%%%%%

\bibliography{./bibli.bib}
%%%%%%%%%%%%%%%%%%%%%%%%%%%%%%%%%%%%%%%%%%%%%%%%%%%%%%%%%%%%%%%%%%%%%%%%%%%%%%%
\clearpage

\newpage

%%%%% END MATTER %%%%%%%%%%%%%%%%%%%%%%%%%%%%%%%%%%%%%%%%%%%%%%%%%%%%%%%%%%%%%%
\section{End Matter}

\subsection{A simplified view of the variations in effective hole-doping}

\begin{figure*}
    \centering
    \includegraphics[width=0.85\linewidth]{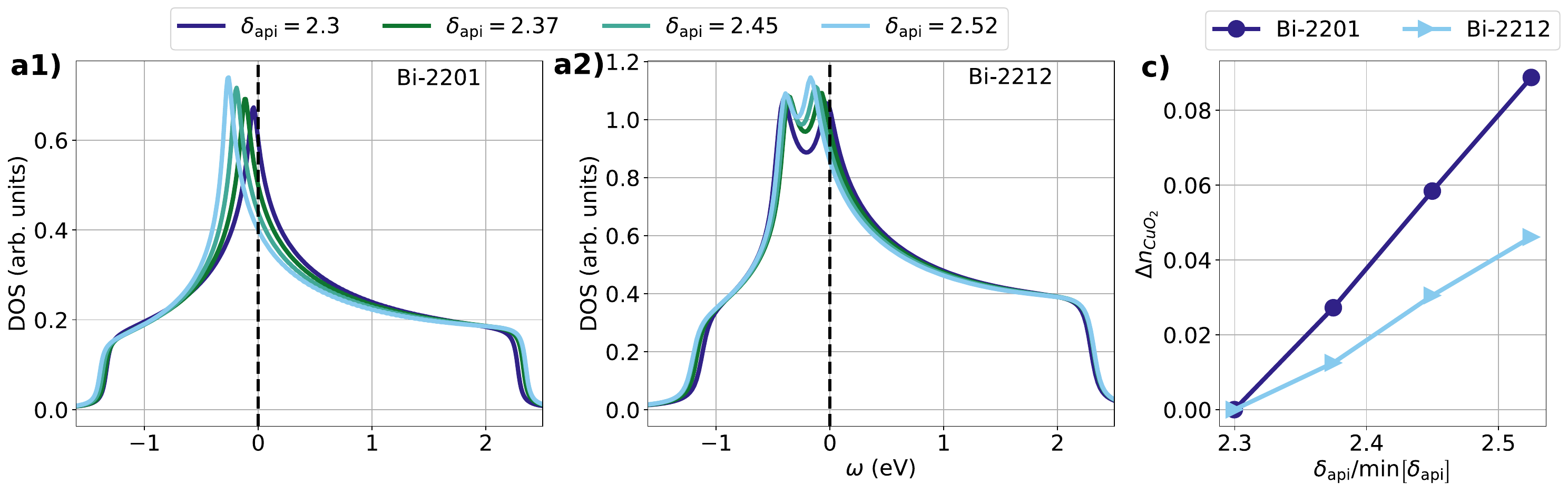}
    \caption{ Non-interacting DOS of the low-energy effective model of (a1) Bi-2201 and (b1) Bi-2212 for all values of $\dapi$. 
    (b) Variations of occupation $\Delta n_{\rm CuO_2}$ estimated from the non-interacting effective models. 
    }
    \label{fig:figem3}
\end{figure*}

In the main text, we interpreted the variations of occupation in terms of charge transfer between different planes and electronic degrees of freedom in the crystal structure. 
While physically transparent, this picture is not easily translated into the language of low-energy effective models, which is commonly used in the cuprate literature.
The purpose of this section is therefore to provide an alternative, simplified description of these effects within the framework of a single-band Hubbard model~\cite{hubbard1963,gutzwiller1963,kanamori1963}.

Our starting point is the DFT-only calculations performed for each apical oxygen distance in Bi-2201/2212. 
We parameterize a single-band Hubbard model using Wannier orbitals: we keep the in-plane hopping terms $t$, $t'$, $t''$, up to the next-next-nearest neighbor, and for Bi-2212 we include the inter-plane hopping term $t_{\perp}$ (001), $t'_{\perp}$ (111) and $t''_{\perp}$ (201/021). 
The corresponding Hamiltonians read: 
\begin{align}
 & H_{2201}(\dapi) = -\sum_{\left<i,j\right>,\sigma}t_{ij}(\dapi)\left[\cdag{i,\sigma}{}\cc{j,\sigma}{}+\mathrm{h.c.}\right]  \\
            & + U\sum_{i}n_{i,\uparrow}n_{i\downarrow} + \left[\epsilon(\dapi)-\mu\right]\sum_{i,\sigma} n_{i,\sigma}, \nonumber \\
 & H_{2212}(\dapi) =  -\sum_{\left<i,j\right>,\sigma,\alpha}t_{ij}(\dapi)\left[\cdag{i,\sigma}{\alpha}\cc{j,\sigma}{\alpha}+\mathrm{h.c.}\right] \\
 & -\sum_{\left<i,j\right>,\sigma,\alpha}t_{\perp,ij}(\dapi)\left[\cdag{i,\sigma}{\alpha}\cc{j,\sigma}{\overline{\alpha}}+\mathrm{h.c.}\right] \nonumber\\
 & + U\sum_{i}n_{i,\uparrow}n_{i\downarrow} + \left[\epsilon(\dapi)-\mu\right]\sum_{i,\sigma} n_{i,\sigma}, \nonumber
\end{align}
where $t_{ij}=t$, $t'$ or $t''$, and $t_{\perp,ij}=t_{\perp}$, $t'_{\perp}$ or $t''_{\perp}$, $\alpha$ and  $\overline{\alpha}$ designate the CuO$_2$ plane in the bi-layer model, $\epsilon$ is the on-site energy, and  $\mu$ the chemical potential.
All model parameters are taken to depend on $\dapi$, except for the chemical potential.
We compute the non-interacting ($U=0$) density of states (DOS) for the effective $H_{2201/2212}$ models at each value of $\dapi$.
The same chemical potential $\mu$ is used for all $\dapi$, reflecting the fact that the remainder of the crystal acts as an effective reservoir that fixes the chemical potential of the CuO$_2$ planes.
We choose $\mu$ such that the $\dapi=2.30~\text{\AA}$ models are effectively at $20\%$ hole-doping, i.e., in a representative doping range compared to our calculations (our analysis is not sensitive to this specific choice of $\mu$). 

The resulting DOS are shown in Fig.~\ref{fig:figem3}(a1,a2). 
For Bi-2201, the particle-hole asymmetry of the DOS increases with $\dapi$. 
As a result, the van Hove singularity shifts towards the occupied part of the spectrum, leading to an increase in the occupation with increasing $\dapi$. 
This is not necessarily related to a change of on-site energy $\epsilon(\dapi)$, but rather to (i) the common chemical potential $\mu$ for all $\dapi$, and (ii) the increase of both $t'/t$ and $t''/t$ with $\dapi$ (we provide the resulting parameters in the SM~\cite{SupMat}). 
Hence, in terms of an effective model, the variations in occupation should not be seen as caused by variations of on-site energies, but rather by variations of particle-hole asymmetry of the underlying non-interacting DOS. 

In the case of Bi-2212, the asymmetry of the non-interacting DOS is also at play, but the mechanism is slightly different. 
We find that the in-plane hoppings are almost independent of $\dapi$, in contrast with the out-of-plane hoppings, which decrease with increasing $\dapi$ (the distance between the two CuO$_2$ planes increases). 
As such, the coupling between the two planes decreases, which leads to a lower splitting between the bonding and anti-bonding bands, as can be seen in Fig.~\ref{fig:figem3}(a2). 
Similarly to Bi-2201, the occupation rises with increasing $\dapi$. 

These results are summarized in Fig.~\ref{fig:figem3}(c).
The increased occupation in the non-interacting effective models for Bi-2201/2212 mimics accurately our DFT+CDMFT results. 
The variations are quantitatively larger than in the realistic calculations,  probably because we ignored the on-site interaction $U$.

\subsection{Witnessing the change in effective hole-doping in the Fermi surface and the spectral function}

\begin{figure}
    \centering
    \includegraphics[width=0.8\linewidth]{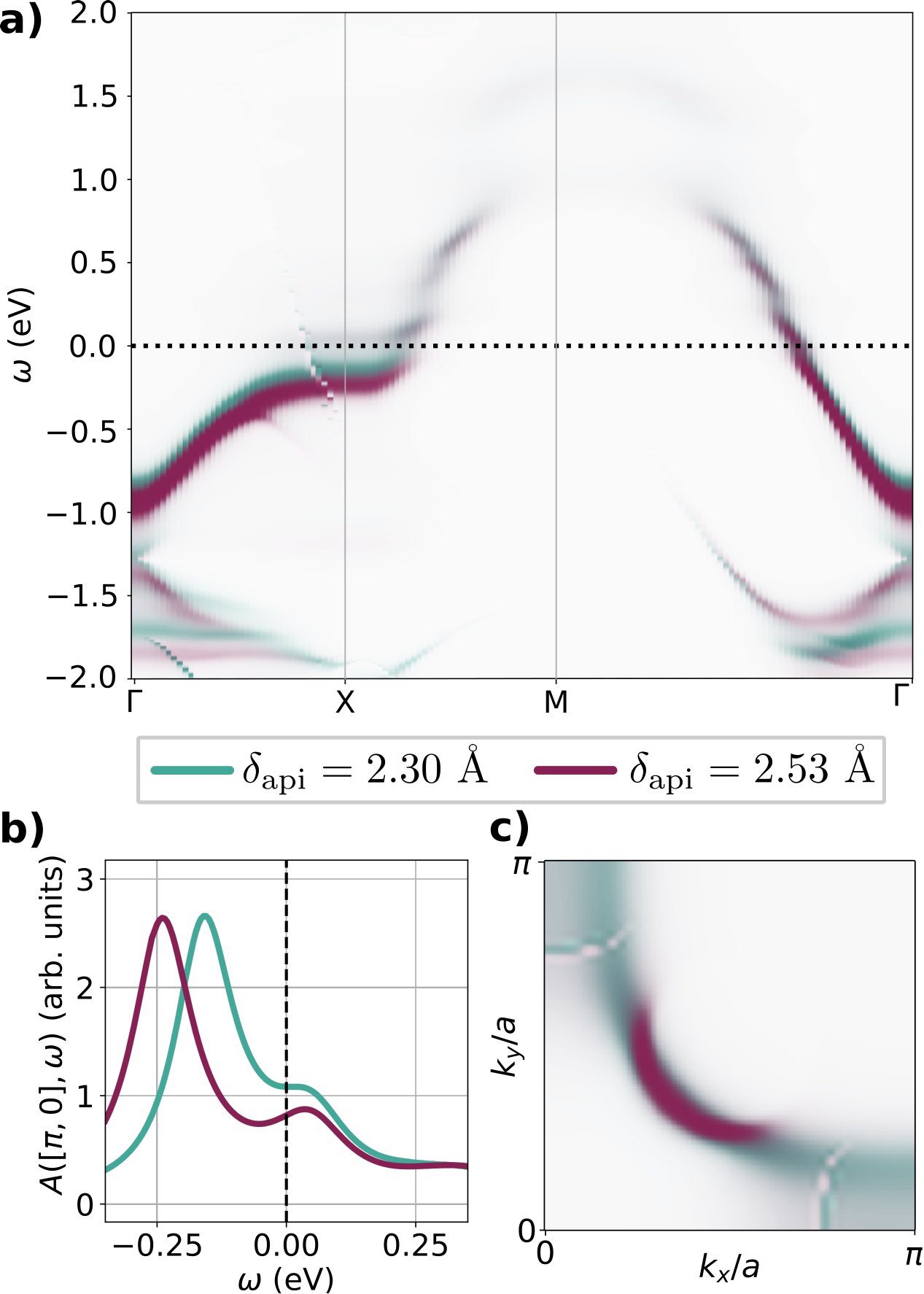}
    \caption{(a) Periodized momentum-resolved spectral function projected on the Cu-$d_{x^2-y^2}$ orbital, in the normal state of Bi-2201 for the two extreme values of $\dapi$. 
    (b) Spectral functions at the anti-nodal $X$ $(\pi,0)$ point. 
    (c) Corresponding super-imposed Fermi surfaces for the two extreme values of $\dapi$. 
    }
    \label{fig:figem2}
\end{figure}

The variations in effective doping of the CuO$_2$ planes are manifest in the spectral properties. 
In Fig.~\ref{fig:figem2}(a) we show the periodized momentum-resolved spectral function of Bi-2201 along the $\Gamma(0,0)-X(\pi,0)-M(\pi,\pi)-\Gamma$ path, for the two extreme apical distances $\dapi=2.30~\text{\AA}$ and $\dapi=2.53~\text{\AA}$.
As expected from the three-band Wannierization detailed in the main text, the two spectral functions are very similar. 
One notices that the occupied spectrum of the smallest apical distance is slightly shifted closer to the Fermi level, consistent with its higher hole-doping. 
The main signature of the change in doping is the pseudogap around the anti-nodal ($X$) region. 
As shown in Fig.~\ref{fig:figem2}(b), the pseudogap is smaller for $\dapi=2.30~\text{\AA}$, since the effective hole doping is larger. 
This finally translates into the Fermi surface shown in Fig.~\ref{fig:figem2}(c). 
The arcs are much more extended for $\dapi=2.30~\text{\AA}$, which again clearly demonstrates that the smallest apical distance corresponds to the largest hole doping.

\subsection{Varying $U$}

We need to check that our results are robust against variations of the on-site interaction $U$. 
Indeed, $U$ is the \emph{only} free parameter in our approach, and it is known to have a substantial effect on the superconducting order parameter $\msc$. 

\begin{figure}
    \centering
    \includegraphics[width=0.85\linewidth]{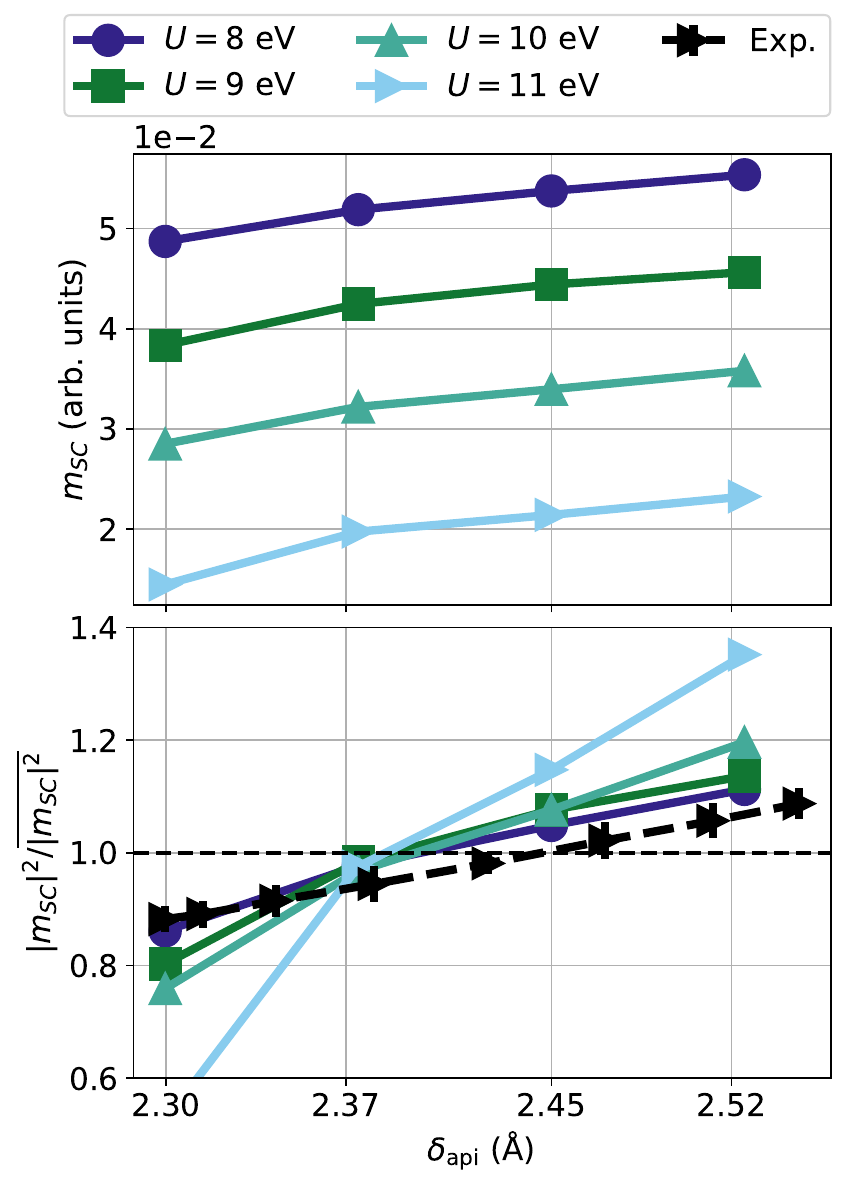}
    \caption{Computed superconducting order parameter $\msc$ vs $\dapi$ for Bi-2212 using different values of $U$ (top).
    Note the $y$-axis scaling.
    Relative variations of the computed and measured superfluid density $|\msc|^2/\overline{|\msc|^2}$ vs $\dapi$ (bottom).
    The experimental data are digitized from Ref.~\onlinecite{omahony_electron_2022}.
    }
    \label{fig:figem1}
\end{figure}

We therefore compute $\msc$ with respect to $\dapi$ in Bi-2212 using a range of on-site interactions $U=8,9,10,11~\text{eV}$. 
As shown in Fig.~\ref{fig:figem1}, the absolute value of $\msc$ is, as expected, inversely correlated with $U$. 
However, the relative variations of the superfluid density $|\msc|^2/\overline{|\msc|^2}$ for $U=8-10~\text{eV}$ remain in quantitative agreement with the STM measurements of Ref.~\cite{omahony_electron_2022}.
Interestingly, a deviation is observed for $U=11~\text{eV}$. 
This is consistent with the fact that self-doping effectively brings the CuO$_2$ into the overdoped regime. 
In this regime, studies of the covalent Emery model show that increasing $U$ pushes the frontier of the superconducting dome to lower critical doping and increases the steepness of the dome~\cite{dash2019,kowalski2021}. 
At $U=11~\text{eV}$, the system has reached the end of the overdoped side of the dome, and thus the relative variations increase more drastically. 

Hence, this additional analysis not only demonstrates the robustness of our results but also reinforces our interpretation based on variations of the effective hole doping. 

%%%%% supplemental material %%%%%%%%%%%%%%%%%%%%%%%%%%%%%%%%%%%%%%%%%%%%%%%%%%%%%%%%%%%%%%
\newpage
\onecolumngrid
\section{supplemental material for "the role of the apical oxygen in cuprate high-temperature superconductors"}

\subsection{Computational details}

\paragraph{Structure optimisation -- }
The study of Bi-based cuprates in their full structural complexity within our density functional theory plus cluster dynamical mean-field theory (DFT+CDMFT) framework poses a considerable challenge that we cannot overcome. 
The primary difficulty stems from the long-range supermodulation of the crystal structure~\cite{levin1994, andersen2007, he2008, zou2020, omahony_electron_2022, jin2024}, which necessitates an enlarged effective unit cell containing several hundred atoms. 
Since our main goal is to isolate the effect of the apical distance $\dapi$, we decompose the Bi$_2$Sr$_2$CuO$_{6+\delta}$ (Bi-2201) supercell into a series of homogeneous tetragonal systems in which $\dapi$ is varied following values expected from experiments~\cite{omahony_electron_2022}.
The resulting structures are subsequently relaxed within DFT + single-site DMFT while maintaining $\dapi$ constant, using the implementation of the eDMFT package~\cite{haule2010,haule2016,haule2018}.

For the structure optimisation of Bi-2201, the correlated subspace contains both Cu-$d_{x^2-y^2}$ and $d_{z^2}$ orbitals.
We use the hybridization-expansion continuous time quantum Monte Carlo solver of Ref.~\onlinecite{haule2007} with $U=12~\text{eV}$ and $J_{\rm Hund}=1~\text{eV}$ at the inverse temperature $\beta=50~\text{eV}^{-1}$, and the exact double-counting (DC) scheme of Ref.~\onlinecite{haule2015} with a mixture of Yukawa and dielectric screening
\begin{equation}
    V_{\rm DMFT}(\mathbf{r},\mathbf{r'}) = \frac{\mathrm{e}^{-\lambda|\mathbf{r}-\mathbf{r'}|}}{\epsilon|\mathbf{r}-\mathbf{r'}|},
\end{equation} 
where $\lambda$ and $\epsilon$ are constants that can be uniquely determined from the value of $U$ and $J_{\rm Hund}$. 
The radius of muffin-tins (RMT) was kept constant for all different homogeneous systems, with their values being given by the most constraining $\dapi$ value.
During this step, a momentum grid of size 12x12x12 is used. 
Force convergence is reached when forces between all atoms (except apical oxygen, which is kept fixed) are smaller than 2.0 mRy/Bohr. 

This structure optimisation step is repeated with various values of unit cell parameters $a$ and $c$ to find the most energetically favorable combination, using splines to interpolate the equilibrium values.
The results of this optimisation are shown in \tref{tab:tab_opti}, where $a$ is found to be constant across all values of $\dapi$ considered, and $c$ varies following $\dapi$ modulations.
With the unit-cell parameters optimised, a final unit-cell is constructed for each $\dapi$ value, and atomic positions are relaxed one last time.

We assumed a similar dependence of $a$ and $c$ with respect to variations of $\dapi$ for the two other compounds -- Bi$_2$Sr$_2$CaCu$_2$O$_{8+x}$ (Bi-2212) and HgBa$_2$CuO$_{4+\delta}$ (Hg-1201).  
The experimental unit cell parameters of Bi-2212 are used for the average value of $\dapi$, and $c$ is adjusted following the $\dapi$ vs $c$ relation of Bi-2201.
The same is done for Hg-1201, but in this case the equilibrium unit-cell corresponds to the largest $\dapi$. 

\begin{table}[h!]
  \begin{center}
    \caption{Unit-cell parameter of the optimised I4/mmm homogeneous Bi-2201 systems}
    \label{tab:tab_opti}
    \begin{tabular}{c||c|c} % <-- Alignments: 1st column left, 2nd middle and 3rd right, with vertical lines in between
      $\delta$ (\AA) & $a$ (\AA) & $c$ (\AA) \\
      \hline \hline
      2.299 & \multirow{4}{*}{3.762} & 23.637\\ 
      2.374 & & 24.130\\ 
      2.449 & & 24.376\\ 
      2.525 & & 24.868\\
    \end{tabular}
  \end{center}
\end{table}

\paragraph{DFT+CDMFT calculations --}
To compute the superconducting order parameter $\msc$, we then resort to DFT+CDMFT calculations. 
We construct $2\times2$ supercells from the optimised structures. 
They can be found in the Open Science Framework repository as \emph{*.cif} files~\cite{OSF-repo}.
To keep the number of atoms tractable in Bi-2201/2212, we first transform the structure from $I4/mmm$ to $P4/mmm$ before constructing the supercell. 
We showed in a previous work that such structure simplification in cuprates leaves the low-energy subspace unchanged~\cite{bacq-labreuil_towards_2025}. 
For each structure, we fix the RMT value for all apical distances and take a $2\times2$ plaquette of Cu-$d_{x^2-y^2}$ orbitals as the correlated subspace. 
We set $U=9~$eV and $J_{\rm Hund}=1~\text{eV}$. 
The impurity problem is solved with exact diagonalization~\cite{PyQCM} using a fictitious inverse temperature $\beta=50~\mathrm{eV}^{-1}$ for the definition of the Matsubara frequencies. 
We use the exact DC scheme as described above. 
Note that a smaller $U$ value is used for cluster calculations compared to our previous study~\cite{bacq-labreuil_towards_2025}. 
This is because the DC correction now includes the angle-dependence of the Cu-$d_{x^2-y^2}$ orbital, which was ignored previously, thereby changing the estimation.  
As discussed in the End Matter, our results are robust against large variations of $U$. 
We use a $12\times12\times8$ k-grid for the DFT+CDMFT convergence in the normal state. 
The virtual crystal approximation (VCA)~\cite{bellaiche_virtual_2000} is applied to the Hg atoms in Hg-1201. 

To compute the order parameter $\msc$, we use the same post-processing method as described in Ref.~\onlinecite{bacq-labreuil_towards_2025}, with a $16\times16\times4$ k-grid.
For Bi-2212, we include proximity effects between the two adjacent CuO$_2$ planes. 
The order parameter $\msc$ is obtained from the lattice average of the pair operator $\hat{D}_{\rm SC}$:
\begin{equation}
\begin{split}
	& \hat{D}_{\mathrm{SC}} = \frac{1}{2}\left[\sum_{\braket{ij}_{x}}\left(d^{\dagger}_{i\uparrow}d^{\dagger}_{j\downarrow}-d^{\dagger}_{i\downarrow}d^{\dagger}_{j\uparrow}\right) - \sum_{\braket{ij}_{y}}\left(d^{\dagger}_{i\uparrow}d^{\dagger}_{j\downarrow}-d^{\dagger}_{i\downarrow}d^{\dagger}_{j\uparrow}\right)+\mathrm{H.c.}\right], \\
	& \msc = \frac{1}{\beta L N_{k}}\mathrm{Tr}_{\omega_n,\mathbf{k}}\left(G(\mathbf{k},i\omega_n)\cdot\hat{D}_{\rm SC}(\mathbf{k})\right),
\end{split}
\end{equation}
where $\mathrm{Tr}_{\omega_n,\mathbf{k}}(.)$ means the trace over Matsubara frequencies $i\omega_n$, superlattice momentum $\mathbf{k}$ and cluster indices, $L$ the number of cluster orbitals and $N_k$ the number of $k$-points. 

A detailed presentation of the method is provided in Ref.~\onlinecite{bacq-labreuil_towards_2025}.

\subsection{Effective Emery--Varma--Schmitt-Rink--Abrahams Hamiltonian}

The Emery--Varma--Schmitt-Rink--Abrahams three-band model~\cite{emery1987,varma1987} reads
\renewcommand\arraystretch{1.5} % for matrix to be a little more readable
\begin{align}
    &\mathcal{H} = \sum_{\vk,\sigma}\Psi^{\dagger}_{\vk,\sigma}\mathbf{h}_0(\vk)\Psi^{\phantom{\dagger}}_{\vk,\sigma} + U\sum_{i}n^{d}_{i\uparrow}n^{d}_{i\downarrow}, \\
     &\mathbf{h}_0(\vk) = \begin{pmatrix}
      - \mu & t_{pd}(1-\expp{-ik_x}) & t_{pd}(1-\expp{-ik_y})\\
     t_{pd}(1-\expp{ik_x}) & \Delta_{p}- \mu & \substack{t_{pp}(1-\expp{ik_x})\\ \cdot(1-\expp{-ik_y})} \\
     t_{pd}(1-\expp{ik_y}) & \substack{t_{pp}(1-\expp{-ik_x})\\\cdot (1-\expp{ik_y})} & \Delta_{p}- \mu
     \end{pmatrix}, \nonumber
\end{align}
\renewcommand\arraystretch{1}% back to normal
where the spinor $\Psi^{\dagger}_{\vk,\sigma}=(d^{\dagger}_{\vk,\sigma},p^{x\dagger}_{\vk,\sigma},p^{y\dagger}_{\vk,\sigma})$ contains the creation operators for the Cu-$d_{x^2-y^2}$ and O-$p_{x/y}$ orbitals at wave vector $\vk$ and spin $\sigma$.
Setting the Cu-$d_{x^2-y^2}$ on-site energy to zero, $\Delta_p$ denotes the on-site energy of the O-$p_{x/y}$ orbitals, $\mu$ is the chemical potential, and $t_{pd}$ ($t_{pp}$) are the Cu--O (nearest O-O) hopping amplitudes.

\subsection{Hopping parameters for the effective single-band models}

In the End Matter, we provided a simplified view of the variations in effective hole-doping in Bi-2201 and Bi-2212 using single-band Hubbard models obtained with Wannier90~\cite{mostofi2008,mostofi2014}.
We present in Table~\ref{tab:tab_hub_Bi2201} and Table~\ref{tab:tab_hub_Bi2212} the parameters for Bi-2201 and Bi-2212, respectively.
The on-site energy $\epsilon$ is rigidly shifted for all apical distances such that the lowest apical distance reaches an effective hole-doping of $20\%$. 

\begin{table}[h]
\centering
\begin{tabular}{c||c|c|c|c}
$\delta_{api}$ ($\mathrm{\AA}$) & $t$ (eV) & $t'/t$ & $t''/t$ & $\epsilon/t$ \\ \hline \hline
2.299  & 0.452 & -0.143 & 0.083 & -0.808 \\ \hline
2.374  & 0.459 & -0.166 & 0.102 & -0.817 \\ \hline
2.449  & 0.465 & -0.184 & 0.117 & -0.786 \\ \hline
2.525  & 0.468 & -0.200 & 0.128 & -0.737 \\ \hline
\end{tabular}
\caption{Hopping parameters $t$, $t'$, $t''$ and on-site energy $\epsilon$ of Bi-2201 single-band Hubbard model.
}
\label{tab:tab_hub_Bi2201}
\end{table}

\begin{table}[h]
\centering
\begin{tabular}{c||c|c|c|c|c|c|c}
$\delta_{api}$ ($\AA$) & $t$ (eV) & $t'/t$ & $t''/t$ & $t_{\perp}/t$ & $t'_{\perp}/t$ & $t''_{\perp}/t$ & $\epsilon/t$ \\ \hline
  2.299 &  0.427 &  -0.246 &  0.149 &  0.166 &  -0.061 &  0.029 & -0.986 \\ \hline
  2.374 &  0.436 &  -0.234 &  0.148 &  0.111 &  -0.045 &  0.020 & -0.967 \\ \hline 
  2.449 &  0.440 &  -0.238 &  0.153 &  0.093 &  -0.045 &  0.015 & -0.933 \\ \hline
  2.525 &  0.441 &  -0.247 &  0.159 &  0.085 &  -0.043 &  0.014 & -0.912 \\ \hline
\end{tabular}
\caption{Hopping parameters $t$, $t'$, $t''$, $t_\perp$, $t'_\perp$, $t''_\perp$ and on-site energy $\epsilon$ of Bi-2212 single-band bi-layer Hubbard model.
The inter-layer hopping $t_{\perp}$ correspond to hopping in the $(0,0,\pm1)$ direction, $t'_{\perp}$ to $(\pm 1,\pm 1,\pm 1)/(\pm 1,\mp 1, \pm 1)$ and $t''_{\perp}$ to $(\pm 2, 0, \pm 1)/(0, \pm 2, \pm 1)$.
}
\label{tab:tab_hub_Bi2212}
\end{table}

\subsection{Hg-1201: investigating further the role of the CTG. }

\begin{figure}
	\centering
	\includegraphics[width=\textwidth]{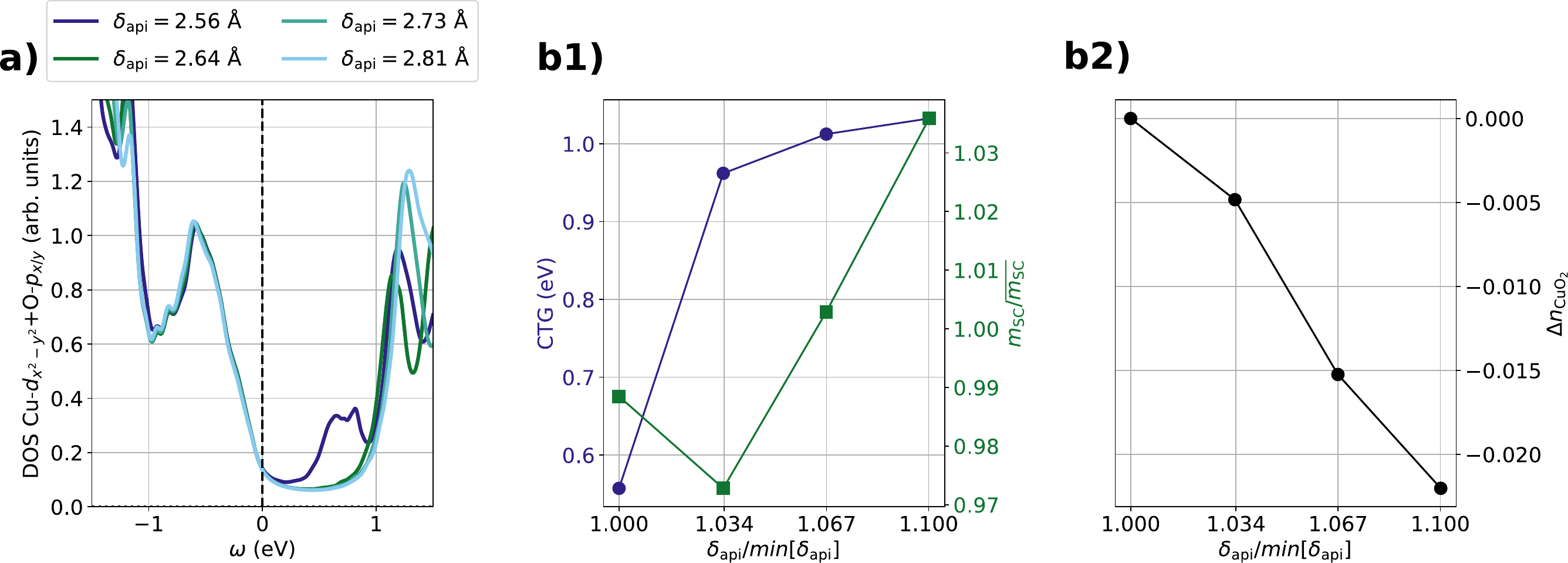}
	\caption{(a) DFT+CDMFT DOS projected on the Cu-$d_{x^2-y^2}$ and O-$p_{x/y}$ orbitals of undoped Hg-1201. 
	(b1) Charge transfer gap (dark blue) extracted from the projected density of states for undoped Hg-1201 with respect to the apical distance. 
	Relative variations of $\msc$ (green) for underdoped Hg-1201 vs. $\dapi$. 
	(b2) Variations of the CuO$_2$ plane occupation vs. the apical distance in underdoped Hg-1201, using the lowest apical distance as reference. 
}
	\label{fig:figS1}
\end{figure}

The Hg-based compound enables further study of the link between the CTG and the apical distance $\dapi$. 
Indeed, in undoped Hg-1201, the CuO$_2$ plane remains insulating, thus granting access within DFT+CDMFT to the well-defined CTG~\cite{bacq-labreuil_towards_2025}.
We performed calculations for undoped Hg-1201 for all apical distances.
The CTG is unambiguously visible in the density of states (DOS) projected on the Cu-$d_{x^2-y^2}$ and O-$p_{x/y}$ orbitals, as shown in Fig.~\ref{fig:figS1}(a). 
As anticipated in the case of Bi-based cuprates, the CTG is \emph{weakly correlated} with $\dapi$. 

The evolution of the CTG, of $\msc$ and of the CuO$_2$ plane occupation $n_{\rm CuO_2}$ vs $\dapi$ are summarized in Fig.~\ref{fig:figS1}(b1,b2).
We recall that the CTG is obtained from undoped Hg-1201, while $\msc$ and $\Delta n_{\rm CuO_2}$ are extracted from calculations in the underdoped regime.
Overall, both $\msc$ and the CTG increase with $\dapi$, while $\Delta n_{\rm CuO_2}$ decreases. 
It appears clearly that the CTG \emph{competes} with the variations of hole doping in Hg-1201: for $\dapi/\mathrm{min}[\dapi]>1.034$ the small increase of CTG (detrimental to $\msc$) is overwhelmed by the rise of the effective hole doping (favorable to $\msc$ in the underdoped regime).
This competition is confirmed by the first decrease of $\msc$ between the two lowest $\dapi$: the small rise in hole doping is overwhelmed by the drastic increase of the CTG. 

This additional analysis of Hg-1201 strikingly confirms the main message of this work: while the CTG is pivotal to superconductivity in cuprates~\cite{weber2012,kowalski2021,bacq-labreuil_towards_2025}, it cannot explain the relation between $\dapi$ and $\msc$.

\subsection{Wannierization of the self-doping bands in Bi-2201 and Bi-2212}

\begin{figure}
	\centering
	\includegraphics[width=\textwidth]{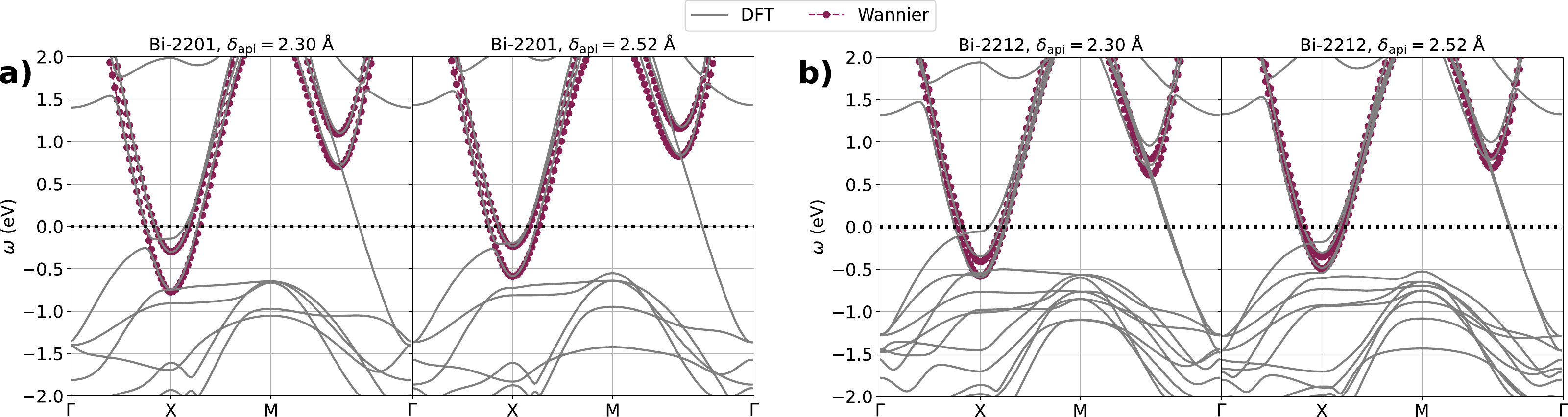}
	\caption{Comparison between the DFT bandstructure and the wannierized self-doping bands in (a) Bi-2201, and (b) Bi-2212.
		For the sake of clarity, the other Wannier bands are not shown.
}
	\label{fig:figS2}
\end{figure}

Increasing the apical distance $\dapi$ leads to opposite variations of the effective occupation of the Cu$-d_{x^2-y^2}$ and O$-p_{x/y}$ orbitals in Bi-2201/2212 ($\Delta n_{\rm CuO_{2}}>0$) and Hg-1201 ($\Delta n_{\rm CuO_{2}}<0$).
We interpret this contrasted behavior in terms of doping channels. 
Both families of compounds share a common channel, related to the global oxydation states, favoring a decrease of the effective occupation upon increasing $\dapi$ (i.e. $\Delta n_{\rm CuO_{2}}$, as in Hg-1201). 
We identified a second channel for Bi-2201/2212: the Bi-O self-doping bands, which rather increases the number of electrons in the CuO$_{2}$ planes.   
The latter is expected to dominate the former, since it involves a trivial shift of two electronic bands at the Fermi level. 
To substantiate our interpretation, we evaluate at the DFT level the expected number of electrons lost in the self-doping Bi-O bands upon increasing $\dapi$, in both Bi-2201 and Bi-2212.  
These electrons are expected to populate the CuO$_{2}$ planes which host the first empty electronic states at the Fermi level. 

Determining the number of electrons trapped in the self-doping bands is not straightforward. 
One cannot rely on projections onto atomic-like orbitals since a substantial part of the spectral weight lies in the interstitial region between the atomic spheres. 
In addition, the self-doping bands are intricated with a few other electronic bands, making their tracking at any $\mathbf{k}$-point difficult. 
To circumvent these difficulties, we wannierized the self-doping bands with Wannier90~\cite{mostofi2008,mostofi2014}, for the lowest and largest $\dapi$ in both Bi-2201 and Bi-2212. 
For Bi-2201, we employed twelve orbitals initialized as the $p_x$ and $p_y$ orbitals of Bi, O$_{\rm api}$ (apical oxygen) and O$_{\rm Bi}$ (oxygen in the BiO planes).
We were able to converge satisfying Wannier bands with only eight orbitals in Bi-2212, using as a starting point the $p_x$ and $p_y$ orbitals of Bi and O$_{\rm Bi}$. 
This procedure provides effective bands reproducing accurately the self-doping bands, and completely disentangled from other bands below the Fermi level, as shown in Fig.~\ref{fig:figS2}. 
Note that we only show the two relevant Wannier bands.
The agreement with the DFT bandstructure is excellent below the Fermi level, i.e. in the region of interest. 

The occupation of the self-doping bands can be obtained trivially with the Wannier bands by computing the corresponding DOS and integrating it up to the Fermi level. 
We computed the DOS with the tetrahedron method implemented in the libtetrabz package~\cite{kawamura2014}, using a $36\times36\times12$ $\mathbf{k}$-grid.

We estimated an occupation of $0.374~\mathrm{e}^{-}$ ($\dapi=2.30~\text{\AA}$) and $0.299~\mathrm{e}^{-}$ ($\dapi=2.52~\text{\AA}$) in the self-doping bands of Bi-2201. 
Hence, we expect that the self-doping channel provides $0.075~\mathrm{e}^{-}$ to the CuO$_2$ planes as $\dapi$ is increased. 
This value is slightly larger than the estimated value from the DFT+CDMFT calculations (roughly $\Delta n_{\rm CuO_{2}} = 0.055~\mathrm{e}^{-}$). 
Interestingly, the difference between the self-doping contribution ($0.075~\mathrm{e}^{-}$) and the observed change ($\Delta n_{\rm CuO_{2}} = 0.055~\mathrm{e}^{-}$) corresponds well to the occupation change in Hg-1201 upon increasing $\dapi$ ($\Delta n_{\rm CuO_{2}} = -0.022~\mathrm{e}^{-}$).
This suggests that the competing global charge transfer channel, common to Hg- and Bi-based cuprates, is indeed competing with the self-doping channel.
As such, directly quantifying the self-doping channel would also provide a good estimate of the other channel. 

For Bi-2212, the occupation of the self-doping band is evaluated at $0.340~\mathrm{e}^{-}$ ($\dapi=2.30~\text{\AA}$) and $0.29~\mathrm{e}^{-}$ ($\dapi=2.52~\text{\AA}$). 
In total, we thus expect an increase of occupation of roughly $0.05~\mathrm{e}^{-}$. 
Since each Bi-2212's unit-cell contains\emph{two} CuO$_{2}$ planes, the expected increase of electron per plane should be $0.025~\mathrm{e}^{-}$. 
Again, this value is slightly larger than the DFT+CDMFT estimate of $0.019~\mathrm{e}^{-}$, but the difference between the two is smaller than in Bi-2201.
This might signal that the opposite doping channel is less affected by $\dapi$ in bi-layer cuprates in contrast to single-layer compounds.

\end{document}